\documentclass[superscriptaddress,twocolumn,aps,pra,showpacs,
fixfloats]{revtex4-1}

\usepackage{bm}
\usepackage{dcolumn}
\usepackage{graphicx}
\begin{document}

\title{Electron elastic scattering off polarizable $A$@C$_{60}$: The complete study within a particular theory}

\author{V. K. Dolmatov}
\affiliation{University of North Alabama,
Florence, Alabama 35632, USA}
\author{ M. Ya. Amusia}
\affiliation{Racah Institute of Physics, Hebrew University, 91904 Jerusalem, Israel}
\affiliation{A. F. Ioffe Physical-Technical Institute, 194021 St. Petersburg, Russia }
\author{L. V. Chernysheva$^{3}$}

 \date{\today}

\begin{abstract}
A deeper insight into electron elastic scattering off endohedral fullerenes $A$@C$_{60}$ is provided. The study accounts for polarization of both the encapsulated atom $A$ and C$_{60}$ cage by an incident electron. It is carried out in the framework of the combination of both a model and the first-principle approximations. A core principle of the  model is that the C$_{60}$ cage itself is presented by an attractive spherical potential of a certain inner radius, thickness, and depth. The main idea of the first-principle approximation is that the polarization of $A$@C$_{60}$ by an incident electron is accounted with the help of   the Dyson equation for the self-energy part of the Green's function of an incident electron. Calculations are performed for, and comparison is made between the individual cases: (a) when $A$@C$_{60}$ is regarded as a static system, (b) when C$_{60}$ is a static cage, but the encapsulated atom is
polarizable, (c) when both C$_{60}$ and $A$ are polarized simultaneously but independently of each other, and (d), as the most general scenario, when polarizabilities of C$_{60}$ and $A$ are coupled.
Spectacular similarities and discrepancies between results of each of the exploited approximations are demonstrated. Revealed features of the overall significant impact of $A$@C$_{60}$ polarization by an incident electron on its elastic scattering off $A$@C$_{60}$ are demonstrated. Choosing Ne, Xe, and Ba as ``probing'' atoms, the dependence of $e + A@{\rm C_{60}}$ scattering on the size and polarizability of the encapsulated atom is unraveled.
\end{abstract}

\pacs{31.15.ap, 31.15.V-, 34.80.Dp, 34.80.Bm}

\maketitle

\section{Introduction}
Electron elastic scattering off quantum targets is an important fundamental phenomenon of
nature. It has significance to both the basic and applied sciences and technologies. To date, however, the knowledge on
electron elastic scattering by such quantum targets as endohedral fullerenes $A$@C$_{60}$, where an atom $A$ is encapsulated
inside the hollow interior of a C$_{60}$ molecule, is rudimentary. This is not accidental, since the comprehensive description
of electron scattering, especially of a low-energy electron scattering,  by a multielectron target  is too challenging
for theorists even with regard to a free atom, not to mention a complicated and more multifaceted $A$@C$_{60}$ target.
It is, perhaps, for this reason that so far there have been only a few attempts undertaken by theorists \cite{DolmJPB,DolmPRA15a,AmusiaJETPL15,DolmPRA15b,AmusiaJETPL16} to advance into
the world  of low-energy electron scattering  off $A$@C$_{60}$. Specifically, to address the problem, a model static Hartree-Fock (HF) approximation was employed in
Refs.~\cite{DolmJPB,DolmPRA15a,AmusiaJETPL15}. Additionally, in Ref.~\cite{AmusiaJETPL15}, a provisional attempt was made to evaluate the impact of polarization of C$_{60}$ on $e + A@{\rm C_{60}}$ scattering.
 Despite a number of interesting features of electron scattering off $A$@C$_{60}$ were predicted, the drawback of those studies was obvious. A further attempt to get a better understanding of $e + A@{\rm C_{60}}$ scattering
was later undertaken in Ref.~\cite{DolmPRA15b}. There,
the model C$_{60}$ cage was kept static (``frozen''), but the encapsulated atom $A$ was ``unfrozen'', i.e., regarded as a polarizable atom. A significant impact of its dynamical polarization (in the presence of static C$_{60}$)
by an incident electron on $e + A@{\rm C_{60}}$ scattering was revealed. That was achieved by exploiting the Dyson formalism \cite{Abrikosov,ATOM} for the self-energy part of the Green's function of a scattered electron.
A remained drawback of work \cite{DolmPRA15b} was the neglect of polarization of C$_{60}$ cage itself by incident electrons. It remained unclear how to overcome this drawback in an efficient and yet physically transparent way. Recently, an idea of how to address the problem of electron scattering off a fully polarizable $A$@C$_{60}$ in a reasonably simple approximation was put forward in Ref.~\cite{AmusiaJETPL16}.

The same idea as in Ref.~\cite{AmusiaJETPL16} is exploited in the present work as well, in order to get a better insight into $e + A@C_{60}$ scattering by accounting for polarization of both C$_{60}$ and $A$. The emphasis is not only on the study of the significance of the polarization effect in $e + A@C_{60}$ scattering, but on the sensitivity of the scattering process to both the size and polarizability of the encapsulated atom $A$. We choose Ne, Xe, and Ba as test-atoms and explore electron elastic scattering off such $A$@C$_{60}$ systems in the region of about a zero to a few eV of the electron energy, where the most interesting effects occur.

Atomic units (a.u.) are used throughout the text unless specified otherwise.

\section{Theoretical concepts}

To give the reader a heads-up on how the problem of $e + A@{\rm C_{60}}$ scattering is going to be addressed in the present work, before indulging into details, the key points to mention are as follows.
As the first step, the C$_{60}$ cage is approximated by an attractive square-well (in the radial coordinate $r$) static potential $U_{\rm c}(r)$ of certain depth $U_{0}$,
inner radius $r_{0}$, and thickness $\Delta$. As a second step, in order to account for the effect of polarization of C$_{60}$ by an incident electron, the C$_{60}$ cage is modeled by a modified potential $V_{{\rm C}_{60}}(r) = U_{\rm c}(r)+V_{\rm s}(r)$. Here, $V_{\rm s}(r)$ is the long-range static polarization potential of C$_{60}$: $V_{\rm s}(r) = -\alpha/[2(r^{2}+b^{2})^{2}]$, where $\alpha$ is the static polarizability of C$_{60}$ and $b$ is a parameter of the order of $r_{0}$. Next,
the wavefunctions of incident electrons and electrons of the encapsulated atom $A$ are calculated in the potential  $V_{A@{\rm C}_{60}}(r)$ which is the sum of $V_{\rm C_{60}}(r)$ and the atomic Hartree-Fock potential
$V_{A}^{\rm HF}(r)$: $V_{A@{\rm C}_{60}}(r) = V_{{\rm C}_{60}}(r)+V_{A}^{\rm HF}(r)$. The thus found functions are substituted into the Dyson's equation for self-energy part of the one-electron Green function $\Sigma$ of an incident electron. In the present work, $\Sigma$ is defined such that it accounts for excitations of both atomic electrons and electrons of the C$_{60}$ cage in the presence of polarized C$_{60}$. Then, the Dyson equation for $\Sigma$ is solved with a reasonable assumption that $r_{\rm p} \gg r_{0} \gg r_{A}$. Here, $r_{\rm p}$ is the electron-projectile distance from the enter of C$_{60}$ and $r_{A}$ is the radius of the encapsulated atom $A$. With the thus determined $\Sigma$, electron elastic-scattering phase shifts upon electron collision with a fully polarizable $A$@C$_{60}$ are found and, eventually, the electron scattering cross sections are calculated, as the last step of the study.

\subsection{Model static HF approximation: ``frozen'' $A@{\rm C}_{60}$}

In the present work, as in Refs.~\cite{DolmJPB,DolmPRA15a,AmusiaJETPL15,DolmPRA15b,AmusiaJETPL16}, the C$_{60}$ cage is modeled by an attractive spherical potential $U_{\rm c}(r)$:
\begin{eqnarray}
U_{\rm c}(r)=\left\{\matrix {
-U_{0}, & \mbox{if $r_{0} \le r \le r_{0}+\Delta$} \nonumber \\
0 & \mbox{otherwise.} } \right.
\label{SWP}
\end{eqnarray}
Here, $r_{0}$, $\Delta$, and $U_{0}$ are the inner radius, thickness, and depth of the potential well, respectively.

The wavefunctions $\psi_{n \ell m_{\ell} m_{s}}({\bm r}, \sigma)=r^{-1}P_{nl}(r)Y_{l m_{\ell}}(\theta, \phi) \chi_{m_{s}}(\sigma)$
and binding energies $\epsilon_{n l}$ of atomic electrons
 are the solutions of a system of the ``endohedral''
HF equations:
\begin{eqnarray}
&&\left[ -\frac{\Delta}{2} - \frac{Z}{r} +U_{\rm c}(r) \right]\psi_{i}
({\bm x}) + \sum_{j=1}^{Z} \int{\frac{\psi^{*}_{j}({\bm x'})}{|{\bm
x}-{\bm x'}|}} \nonumber \\
 && \times[\psi_{j}({\bm x'})\psi_{i}({\bm x})
- \psi_{i}({\bm x'})\psi_{j}({\bm x})]d {\bm x'} =
\epsilon_{i}\psi_{i}({\bm x}).
\label{eqHF}
\end{eqnarray}
Here, $n$, $\ell$,  $m_{\ell}$ and $m_{s}$ is the standard set of quantum numbers of an electron in a central field, $\sigma$ is the electron spin variable, $Z$ is the nuclear charge of the atom, ${\bm x} \equiv ({\bm r}, \sigma)$, and the integration over ${\bm x}$ implies both the integration over ${\bm r}$ and summation over
$\sigma$. Eq.~(\ref{eqHF}) differs from the ordinary HF equation for a free atom by the presence of the $U_{\rm c}(r)$ potential in there. This equation is first solved in order to calculate the electronic ground-state wavefunctions of the encapsulated atom. Once the electronic ground-state wavefunctions are determined, they are plugged back into
 Eq.~(\ref{eqHF}) in place of $\psi_{j}({\bm x'})$ and $\psi_{j}({\bm x})$ in order to calculate the electronic wavefunctions of scattering-states $\psi_{i}({\bm x})$ and their radial parts
 $P_{\epsilon\ell}(r) \equiv P_{\epsilon_{i}\ell_{i}}(r)$ in Eq.~(\ref{eqHF}).

 Corresponding electron elastic-scattering phase shifts $\delta_{\ell}(k)$
  are then determined by
 referring to $P_{k\ell}(r)$ at large $r$:
\begin{eqnarray}
P_{k\ell}(r) \rightarrow \sqrt{\frac{2}{\pi}}\sin\left(k r -\frac{\pi\ell}{2}+\delta_{\ell}(k)\right).
\label{P(r)}
\end{eqnarray}
Here, $P_{k\ell}(r)$ is normalized to $\delta(k-k')$, where $k$ and $k'$ are the wavenumbers of the incident and scattered electrons, respectively.
The total electron elastic-scattering cross section $\sigma_{\rm el}(\epsilon)$ is then found in accordance with the well-known
formula for electron scattering by a central-potential field:
 \begin{eqnarray}
 \sigma_{\rm el}(k)= \frac{4\pi}{k^2}\sum^{\infty}_{\ell=0}(2\ell+1)\sin^{2}\delta_{\ell}(k).
 \label{sigma}
 \end{eqnarray}

This approach solves the problem of $e + A@{\rm C_{60}}$ scattering in a static approximation, i.e.,
without account for polarization of the $A$@C$_{60}$ system by incident electrons.

In the literature, some inconsistency is present in
choosing the magnitudes of $\Delta$, $U_{0}$ and $r_{0}$ of the model potential $U_{\rm c}(r)$ for C$_{60}$, for example:
$r_{0}=5.8$, $\Delta=1.9$ and $U_{0}=0.302$~a.u.\ \cite{JPCVKDSTM99,O'Sullivan13} (and references therein),
or $r_{0}= 6.01$, $\Delta=1.25$ and $U_{0}=0.422$~a.u.\ \cite{DolmJPCS12,Gorczyca12}, or   $r_{0} = 5.262$, $\Delta = 2.9102$, and $U_{0} = 0.2599$~a.u.\
\cite{DolmJPB,DolmPRA15a,AmusiaJETPL15,DolmPRA15b,AmusiaJETPL16,Winstead} (originally suggested in \cite{Winstead}).
A better choice of the parameters with an eye on $e + {\rm C_{60}}$ scattering
was investigated in Refs.~\cite{DolmPRA15b,DolmJPCS15}. The made conclusion was in favor of the latter set of parameters: $r_{0} = 5.262$, $\Delta = 2.9102$, and $U_{0} = 0.2599$~a.u.
This is because the chosen set of parameters leads to a better agreement between some of the most prominent features of
$e + {\rm C_{60}}$ elastic scattering predicted by the described model and the sophisticated \textit{ ab initio} static multiconfigurational Hartree-Fock approximation \cite{Winstead}.
Correspondingly, in the present work,
$U_{\rm c}(r)$ potential is defined by
$\Delta = 2.9102$, $r_{0} = 5.262$, and $U_{0} = 0.2599$~a.u.

\subsection{Polarizable $A@{\rm C}_{60}$}
\subsubsection{Polarizable $A$ but ``frozen'' ${\rm C}_{60}$}

Let us first account for the impact of dynamical polarization
of only an atom $A$, encapsulated inside \textit{frozen} C$_{60}$, on $e + A@{\rm C}_{60}$
elastic scattering. This should help, later in the paper, to appreciate the importance of the effect of  C$_{60}$ polarization, in addition to that of $A$,
on $e + A@{\rm C}_{60}$ scattering, to finalize the study.

\begin{figure}[h]
\includegraphics[width=\columnwidth]{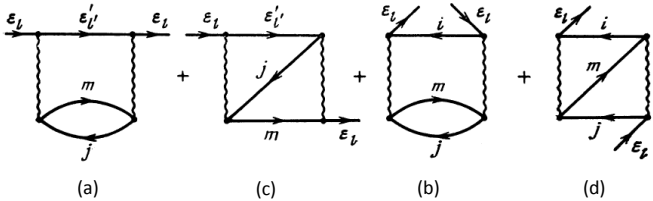}
\caption{The irreducible self-energy part $\Sigma(\epsilon)$
of the Green function of a
scattering electron in the second-order perturbation theory in the Coulomb interaction, referred to as the SHIFT approximation (see text). Here,
a line with a right directed arrowhead denotes either scattered states $|\epsilon_{\ell}\rangle$ and $|\epsilon_{\ell^{\prime}}^{\prime}\rangle$, or atomic excited states $|m \rangle$, a line with a
left directed arrowhead denotes the states $\langle j|$ and $\langle i|$ of a vacancy (hole) in the atom , and a
wavy line denotes the Coulomb interelectron interaction $V$.}
\label{SHIFT}
\end{figure}

In the present work, as in Ref.~\cite{DolmPRA15b}, the impact of polarization of $A$ on $e + A@{\rm C}_{60}$ scattering (where the C$_{60}$ cage is ``frozen'') is accounted on the first-principle basis by utilizing the concept of the self-energy part of the
 Green's function of an incident electron \cite{Abrikosov,ATOM}.
In the simplest second-order perturbation theory in the Coulomb
interelectron interaction $V$ between the incident and atomic electrons,
the  \textit{irreducible} self-energy part of the
Green's function $\Sigma(\epsilon)$ of the  incident  electron
 is depicted with the help of Feynman diagrams in Fig.~\ref{SHIFT}.
%

The diagrams of Fig.~\ref{SHIFT} illustrate how a scattered electron ``$\epsilon _{\ell }$'' polarizes a $j$-subshell
of the atom by causing $j$ $\rightarrow $ $m$ excitations from the subshell and couples with
these excited states itself via both the Coulomb direct [diagrams (a) and (b)] and exchange [diagrams (c) and (d)] interactions.
Numerical calculations of electron elastic-scattering phase shifts, in the framework of this approximation, are performed with the help of the computer code from Ref.~\cite{ATOM} labeled as ``SHIFT''.
Correspondingly, the authors refer to this approximation as the ``SHIFT'' approximation everywhere in the present paper.

\begin{figure}[h]
\includegraphics[width=\columnwidth]{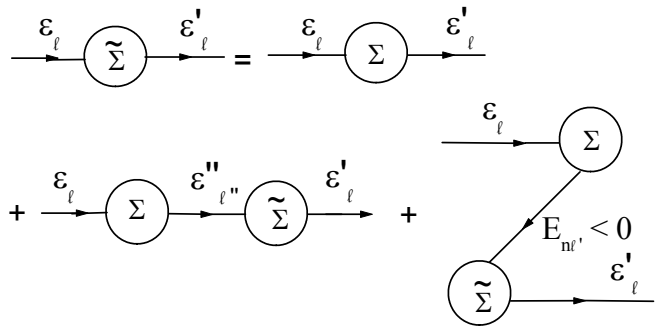}
\caption{The matrix element of the reducible self-energy part $\tilde{\Sigma}(\epsilon)$
of the Green's function of a scattering electron, where $\Sigma$ is the irreducible self-energy part of the Green's function depicted in Fig.~\ref{SHIFT}. This approximation is
referred to as the SCAT approximation (see text). Note, when calculating
$\langle \epsilon_{\ell}|\tilde{\Sigma}|\epsilon_{\ell}\rangle$ analytically, the summation over unoccupied discreet and integration over continuum excited states (marked as $\epsilon_{\ell ''}''$)
along with the summation over the occupied states in the atom marked as $E_{n\ell '}$ must be performed.}
\label{SCAT}
\end{figure}

A fuller account of the effect of the encapsulated-atom polarization in $e + A@{\rm C}_{60}$ elastic scattering is determined by the \textit{reducible}
$\tilde{\Sigma}(\epsilon)$ part of the self-energy part of
the electron's Green function \cite{ATOM}. The matrix element of the latter are represented diagrammatically in Fig.~\ref{SCAT}.
%

The above diagrammatic equation for $\tilde{\Sigma}$ can be written in an operator form \cite{ATOM}, as follows:
\begin{eqnarray}
\hat{\tilde{\Sigma}}=\hat{\Sigma}-\hat{\Sigma}\hat{G}^{(0)}\hat{\tilde{\Sigma}}
\label{EqGreen}.
\end{eqnarray}
Here, $\hat{\Sigma}$
is the operator of the  irreducible
self-energy part of the Green's function calculated in the
framework of SHIFT (Fig.~\ref{SHIFT}), $\hat{G}^{(0)}=(\hat{H}^{(0)}-\epsilon)^{-1}$ is the electron Green's function provided $\hat{H}^{(0)}$ is the
Hamiltonian of an incident electron in a  HF approximation (in the presence of the C$_{60}$ confinement). Clearly, the equation for  $\tilde{\Sigma}$
accounts for an infinite series of diagrams by coupling the diagrams of Fig.~\ref{SHIFT} in various combinations.
Numerical calculation of electron elastic-scattering phase shifts in the framework of this approximation is performed with the help of the computer code from Ref.~\cite{ATOM} labeled as ``SCAT''.
Correspondingly, this approximation is referred to as ``SCAT'' everywhere in the present paper.
SCAT works well for the case of electron scattering off free atoms \cite{ATOM}.
This gives us confidence in that SCAT is a sufficient approximation for pinpointing the impact of correlation/polarization
of $A$@C$_{60}$ electrons on $e + A@{\rm C_{60}}$ scattering as well.

In the framework of SHIFT or SCAT, the electron elastic-scattering
phase shifts $\zeta_{\ell}$  are determined as follows \cite{ATOM}:
\begin{eqnarray}
\zeta_{\ell }=\delta_{\ell }^{\rm HF}+\Delta\delta_{\ell }.
\end{eqnarray}
Here, $\Delta\delta_{\ell}$ is the correlation/polarization
correction term to the calculated HF phase shift $\delta_{\ell}^{\rm HF}$ \cite{ATOM}:
\begin{eqnarray}
\Delta\delta_{\ell}=\tan^{-1}\left(-\pi
\left\langle\epsilon\ell|\tilde{\Sigma}|\epsilon\ell\right\rangle \right).
\label{Delta}
\end{eqnarray}
The mathematical expression for
$\left\langle\epsilon\ell|\tilde{\Sigma}|\epsilon\ell\right\rangle$
is cumbersome. The
interested reader is referred to \cite{ATOM} for details. The matrix element
$\left\langle\epsilon\ell|\tilde{\Sigma}|\epsilon\ell\right\rangle$ becomes
complex for electron energies exceeding the
ionization potential of the atom-scatterer, and so does the correlation term
$\Delta\delta_{\ell}$ and, thus, the phase shift
 $\zeta_{\ell}$ as well. Correspondingly,
\begin{eqnarray}
\zeta_{\ell }=\delta_{\ell}+i\mu_{\ell},
\end{eqnarray}
where
\begin{eqnarray}
\delta_{\ell}=\delta_{\ell }^{\rm HF}+
\rm Re\Delta\delta_{\ell},\quad \mu_{\ell} =
Im\Delta\delta_{\ell }.
\end{eqnarray}

The total electron elastic-scattering cross section $\sigma_{\rm el}$ is then given by the expression
\begin{eqnarray}
\sigma_{\rm el}=\sum_{\ell =0}^{\infty }\sigma_{\ell},
\end{eqnarray}
where $\sigma_{\ell}$ is the electron elastic-scattering partial cross section \cite{ATOM}:
\begin{eqnarray}
\sigma_{\ell}=\frac{2\pi }{k^2}(2\ell+1)(\cosh{2\mu_{\ell}}-
\cos{2\delta_{\ell}}){\rm e}^{-2\mu_{\ell}}.
\label{EqSgmRPAE}
\end{eqnarray}

\subsubsection{Polarizable $A$ and ${\rm C_{60}}$: approximation of ``uncoupled'' polarizabilities}

We now account for the effects of polarization of C$_{60}$ by an incoming electron in addition to the encapsulated atom polarization.
This is done in a simple approximate way, as is often done in atomic physics  \cite{Drukarev}. Namely, the long-range polarization potential of C$_{60}$
is approximated by a static dipole polarization potential $V_{\rm p}(r)$:
\begin{eqnarray}
V_{\rm p}(r) = -\frac{\alpha}{2(r^{2}+b^{2})^{2}}.
\label{EqVs}
\end{eqnarray}
Here, $\alpha$ is the static dipole polarizability of C$_{60}$ and $b$ is the ``cut-off'' parameter of the order of the radius of C$_{60}$, $r_{0}$.

In the present work, the C$_{60}$ effective potential $V_{{\rm C}_{60}}^{\rm eff}(r)$, ``felt'' by an incident electron,
is approximated by the sum of the short-range potential $U_{\rm c}(r)$, Eq.~(\ref{SWP}), and the long-range polarization potential $V_{\rm p}(r)$, Eq.~(\ref{EqVs}):
\begin{eqnarray}
V_{{\rm C}_{60}}^{\rm eff}(r) =U_{\rm c}(r)+V_{\rm p}(r).
\label{Veff}
\end{eqnarray}

In the given approximation, the wavefunctions of the ground and excited states of the atom $A$ are calculated with the help of Eq.~(\ref{eqHF}), where the potential
$U_{\rm c}(r)$ is replaced by $V_{{\rm C}_{60}}^{\rm eff}(r)$.  Such defined wavefunctions are used in calculations
 of  $e + A@{\rm C}_{60}$ elastic-scattering phase shifts with  account for
dynamical polarization of the atom $A$ in the presence of polarizable C$_{60}$. This is done in the framework of the above described Dyson formalism for the self-energy part of the Green function
of an incident electrons, Eqs.~(\ref{EqGreen})-(\ref{EqSgmRPAE}). The determined phase shifts are plugged into Eqs.~(\ref{EqGreen}) and (\ref{Delta}) to meet the goal of the study, that is
to account for the impact of both \textit{dipole static} polarization of C$_{60}$ (in the model approximation)
and \textit{dynamical} polarization of the encapsulated atom $A$ (on the first-principle basis) on  $e + A@{\rm C}_{60}$ scattering.

In the developed approximation, the polarizabilities of C$_{60}$ and $A$
are, obviously, not coupled with each other (no interelectron interaction between the electrons of $A$ and C$_{60}$). This approximation is referred to as the \textit{approximation of ``uncoupled'' polarizabilities} of C$_{60}$ and $A$ in the present paper.

\subsubsection{Polarizable $A$ and C$_{60}$: approximation of ``coupled'' polarizabilities}

\begin{figure}[h]
\includegraphics[width=\columnwidth]{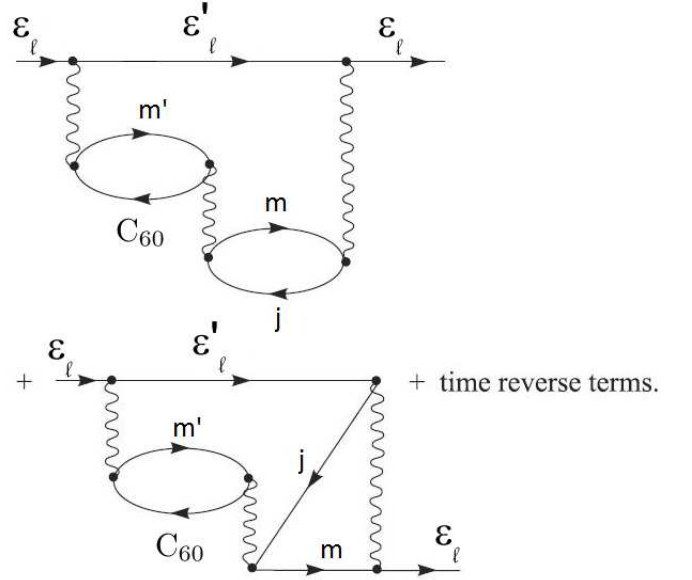}
\caption{The multielectron processes accounting for the interaction between excited configurations of C$_{60}$ and $A$ brought about by an incident electron.
Lines marked as ``C$_{60}$''and ``$m'$'' denote the core and excited stated of the fullerene cage C$_{60}$, respectively. Other notations are the same as in Fig.~\ref{SHIFT}.}
\label{FigFA}
\end{figure}

In the present paper, the term ``\textit{coupled polarizabilities}'' of $A$ and C$_{60}$ means coupling between excited configurations of $A$ and C$_{60}$, both of which are caused by an incident electron. The
corresponding correction to $e + A@{\rm C}_{60}$ scattering is in addition to the above discussed effect of ``uncoupled'' polarizabilities of $A$ and C$_{60}$.
In the present work, this is done, first, by adding some specific third order terms of perturbation theory in Coulomb interaction to the \textit{irreducible} $\Sigma(\epsilon)$ (Fig.~\ref{SHIFT}). Then, thus re-defined $\Sigma(\epsilon)$ is substituted into the equation for the \textit{reducible} part of the Green function $\tilde{\Sigma}(\epsilon)$ [Eq.~(\ref{EqGreen})
and/or Fig.~\ref{SCAT}].  The third order terms in question are depicted with the help of Feynman diagrams in Fig.~\ref{FigFA}.

Strictly speaking, to calculate the terms depicted in Fig.~\ref{SCAT}, one needs to calculate wavefunctions of the ground and excited states of the multielectron fullerene cage C$_{60}$.
In the present paper, we bypass this difficulty by employing  a simple and yet reasonable  approximation. It is consists in the following. First,
it takes into account that the radius $r_{0}$ of the C$_{60}$ cage is bigger than the radius
 $r_{\rm A}$ of an encapsulated atom $A$: $r_{0} > r_{\rm A}$. Second, it exploits the fact that the impact of the polarization potential of C$_{60}$ on electron scattering is essential only at large distances
$r_{\rm p}$ from the target. Hence, $r_{\rm p} > r_{0}$, $r_{\rm p}$ being the important electron-projectile distance from the center of C$_{60}$. Next, as a not too strong exaggeration, let us regard that
$r_{\rm p} \gg r_{0} \gg r_{\rm A}$. Then, the
Coulomb interaction between the incident electron and C$_{60}$'s electrons, $V(\bm{r}_{\rm p}, \bm{r}_{\rm{C}}) \propto |\bm{r}_{\rm p} - \bm{r}_{\rm{C}}|^{-1}$, and that one between the C$_{60}$'s electrons and electrons of the encapsulated atom $A$, $V(\bm{r}_{\rm A}, \bm{r}_{\rm{C}}) \propto |\bm{r}_{\rm A} - \bm{r}_{\rm{C}}|^{-1}$, turn, approximately, into the long-range dipole potentials:
\begin{eqnarray}
V(\bm{r}_{\rm p}, \bm{r}_{\rm{C}}) \propto \frac
{
\bm{r}_{\rm p}\bm{r}_{\rm{C}}
}
{
\bm{r}_{\rm p}^3
}, \quad
V(\bm{r}_{\rm A}, \bm{r}_{\rm{C}}) \propto \frac
{
\bm{r}_{\rm A}\bm{r}_{\rm{C}}
}
{
\bm{r}_{\rm C}^3
}.
\label{ApproxV}
\end{eqnarray}
Hence, the exact Coulomb interactions, represented by the first and second wavy lines in the diagrams of Fig.~\ref{FigFA}, can now be replaced by their approximate values defined by
Eqs.~(\ref{ApproxV}). This is equivalent to accounting for  the impact of only dipole polarization of C$_{60}$ on electron scattering. The impact can be expressed via the
known dipole polarizability $\alpha$ of C$_{60}$, similar to how it has been done in Ref.~\cite{AmBaltPLA06}. One can show that this is equivalent to making simple replacements
\begin{eqnarray}
|V|^{2} \rightarrow |V[1 - \alpha(\epsilon_{\ell} - \epsilon'_{\ell})/r_{0}^{3}]|^{2}
\label{EqV1}
\end{eqnarray}
in the diagrams (a) and (c) of Fig.~\ref{SHIFT}, and
\begin{eqnarray}
|V|^{2} \rightarrow |V[1 - \alpha(\epsilon_{\ell} - \epsilon'_{\ell})/r_{0}^{3}] \nonumber \\
\times V[1 - \alpha(\epsilon_{\ell} - \epsilon_{m} )/r_{0}^{3}]|.
\label{EqV1'}
\end{eqnarray}
in the diagrams (c) and (d) of Fig.~\ref{SHIFT}. Next, the re-calculated  $\Sigma(\epsilon)$ (Fig.~\ref{SHIFT}) is
substituted into Eq.~(\ref{EqGreen}) for $\tilde{\Sigma}(\epsilon)$. The latter concludes the problem of  $ e + A@{\rm C_{60}}$ scattering
in the framework of the approximation of coupled polarizabilities of C$_{60}$ and $A$.

\section{Results and Discussion}

\subsection{Preliminary notes}

First, in order the modeling of the C$_{60}$ cage by the homogeneous spherical potential $U_{\rm c}(r)$, Eq.~(\ref{SWP}), made sense,  the wavelength $\lambda$ of an incident electron
must exceed greatly the bond length $D \approx 1.44$ $\AA$ between the carbon atoms in
C$_{60}$. In the present work, therefore, the maximum energy of an incident electron is capped at approximately $4$ eV. Up to this energy, $\lambda \agt 6$ $\AA$, so that, indeed,
$\lambda \gg D$.

Second, given that the outer radius of C$_{60}$ is approximately $8$~a.u., one can easily estimate that the maximum contribution to electron scattering comes from the first five
electron partial waves with $\ell$ ranging from $0$ to $4$. Correspondingly, in the present work, the maximum value of $\ell$ is capped at $\ell = 4$.

Third, when accounting for multipolar excitations of the electronic subshells of an encapsulated atom $A$ by an incident electron, the performed calculations accounted for  the monopole, dipole, quadrupole, and octupole excitations of outer subshells of the atoms. For the chosen in the present paper set of encapsulated atoms, these are the $6s^{2}$ and $5p^{6}$ subshells in Ba, $5p^{6}$ in Xe, and $2p^{6}$ in Ne.

Fourth, since one of the main goals of this study is to explore how both
the size of an atom and its dynamical polarizability might affect electron scattering off polarizable $A@{\rm C_{60}}$, the study is run
along the path from the most compact to the most diffuse encapsulated atom: $e + {\rm Ne@C_{60}}$ $\rightarrow$ $e + {\rm Xe@C_{60}}$ $\rightarrow$  $e + {\rm Ba@C_{60}}$.

Finally, in order to facilitate the reader to cope with various abbreviations for the utilized approximations, adopted in the present paper,  these are as follows:

\begin{itemize}
\item HF stands for the model static HF approximation, where both the encapsulated atom $A$ and
the fullerene cage C$_{60}$ are regarded as non-polarizable targets (see subsection \textbf{A} above).

\item SCAT(A@) marks the approximation, which accounts for polarization of only the encapsulated atom $A$ (see subsection $1$ of subsection \textbf{B}).

\item SCAT(A) designates the approximation as the above one, but in relation to a free atom $A$.

\item SCAT(UP) labels the approximation of uncoupled polarizabilities of  C$_{60}$ and $A$ (see subsection $2$ of subsection \textbf{B}).

\item SCAT(CP) stands for the approximation of coupled polarizabilities of C$_{60}$ and $A$ (see subsection $3$ of subsection \textbf{B}).

\end{itemize}

\subsection{$e + {\rm Ne@C_{60}}$ scattering}

Based on our previous HF calculated data for $e + A@{\rm C_{60}}$ scattering \cite{DolmJPB,DolmPRA15a,AmusiaJETPL15}, it is known that the encapsulation of Xe into the C$_{60}$ cage practically
does not change electron elastic scattering off $e + A@{\rm C_{60}}$ compared to scattering by empty C$_{60}$ calculated within the model static approximation. This is because the Xe atom
 is compact. Obviously, one can expect the same for even smaller compact atoms, like the Ne atom. In addition, the Ne atom has
a smaller polarizability $\alpha$ than that of Xe or Ba. For example, experimental values of the static dipole polarizabilities of the ground states of Ne, Xe, and Ba are as follows:
 $\alpha({\rm Ne}) \approx 2.67$, $\alpha({\rm Xe}) \approx 27.34$, and $\alpha({\rm Ba}) \approx 269$~a.u.\ \cite{Safronova2010}. Therefore, the study of $e + {\rm Ne@C_{60}}$ scattering makes a particular sense in two respects. First, it allows to understand better how
accounting for polarization of only C$_{60}$ affects electron elastic scattering off C$_{60}$ doped with a compact static atom. Second, the study  can shed light on whether the coupling between the \textit{weak} polarizability  of a compact atom $A$ with the \textit{large} polarizability of C$_{60}$ ($\alpha \approx 850$~a.u.\ \cite{AmBaltPLA06}) can have a noticeable effect on $e + A@{\rm C_{60}}$ scattering compared to that calculated in the SCAT(UP) approximation. Common sense suggests that the effect should be negligible.
However, as one of key findings of the present work, such intuitive perception turns out to be not entirely correct.

\begin{figure}
\center{\includegraphics[width=\columnwidth]{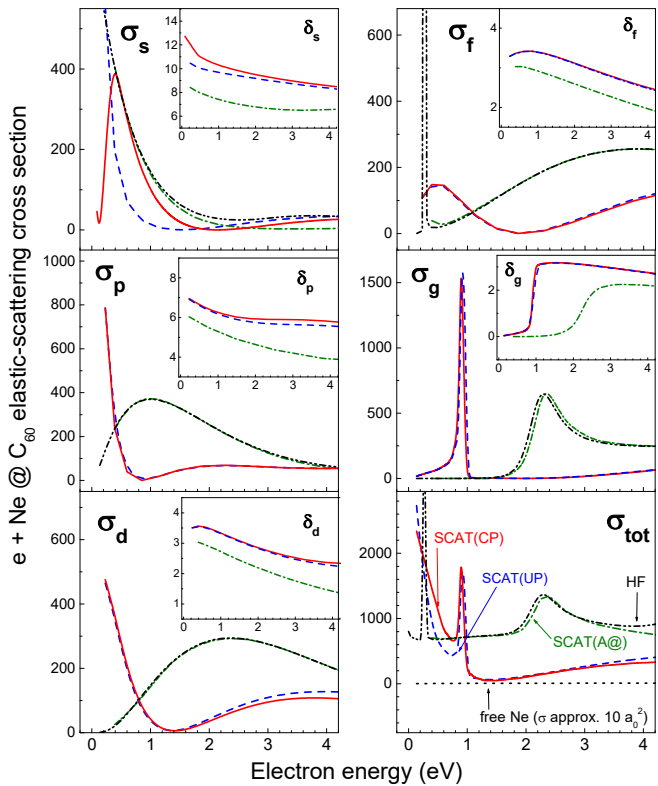}}
\caption{(Color online) Main panels: Calculated partial $\sigma_{\ell}(\epsilon)$ and total $\sigma_{\rm el}(\epsilon)$ cross sections (a.u.)
for electron elastic scattering off Ne@C$_{60}$. Insets: Real parts $\delta_{\ell}(\epsilon)$ of the phase shifts (in units of radian); imaginary parts $\mu_{\ell}=0$ in this energy region. The used styles of the plotted lines mark results obtained
in different approximation utilized in the present work, as follows:  dash-dot-dot, HF; dash-dot, SCAT(A@); dots, SCAT(A) (free Ne); dash, SCAT(UP); solid, SCAT(CP) (the most complete approximation).
Note, the calculated SCAT(A@) and HF data for $p$-, $d$-, and, for a better part of $f$-partial cross sections are so close to each other that one can hardly tell one from the other.}
\label{Ne}
\end{figure}

Calculated HF, SCAT(A@), SCAT(A), SCAT(UP), and SCAT(CP) data for the real parts $\delta_{\ell}(\epsilon)$ of phase shifts, as well as the  partial $\sigma_{\ell}(\epsilon)$ and total $\sigma_{\rm el}(\epsilon)$  cross sections of  a $e + {\rm Ne@C_{60}}$ elastic collision are displayed
in Fig.~\ref{Ne}.
%

One can see that the difference between calculated model static HF data and SCAT(@A) data for all partial electronic waves is negligible. This agrees with common sense, in view of insignificant polarizability of Ne. The data also demonstrate that the polarizability of a free small-sized compact atom, like Ne, remains small upon its encapsulation inside C$_{60}$. Indeed, in the opposite case, the difference between calculated static HF and SCAT(A@) data would have been noticeable.

Let us now explore the calculated data for the $p$-, $d$-, $f$-, and $g$-partial electronic waves scattered off  ${\rm Ne@C_{60}}$. One can conclude that simultaneous accounting for polarization of both Ne and C$_{60}$ by the
scattered electron [SCAT(UP) and SCAT(CP) approximations; dash and solid lines, respectively] has a huge effect on the corresponding cross sections compared to when only polarization of Ne is accounted [SCAT(A@), dash-dot line]. This
was expected in view of a large polarizability of C$_{60}$. Yet, details of how the polarization of C$_{60}$ affects $e + A@{\rm C_{60}}$ scattering have been cleared up only in the present study, to the authors' best knowledge. Next, one can also see that the effect of coupling between the polarizabilities of Ne and C$_{60}$ [SCAT(CP) approximation, solid line] on scattering of these partial electronic waves off $e + {\rm Ne@C_{60}}$ is insignificant. Indeed, the calculated SCAT(CP)  and SCAT(UP) (dash line) data are practically identical. This, again, fits common sense, in view of a small polarizability of Ne.

However, the study finds that $s$-scattering behaves quite differently compared to the above case of scattering of waves with higher $\ell$s. In contrast to the above case, we found
 both big quantitative and qualitative differences between calculated  $\sigma_{s}^{\rm SCAT(UP)}$ (dash line)
and $\sigma_{s}^{\rm SCAT(CP)}$ (solid line). Indeed, $\sigma_{s}^{\rm SCAT(UP)}$ is seen to monotonically increase, whereas, in contrast, $\sigma_{s}^{\rm SCAT(CP)}$ develops a strong maximum
with decreasing electron energy $\epsilon$. The reason for the different behavior of $\sigma_{s}^{\rm SCAT(CP)}$ versus $\sigma_{s}^{\rm SCAT(UP)}$ becomes clear upon exploring the corresponding phase shifts.
One can see that the phase shift $\delta_{s}^{\rm SCAT(CP)}$ (solid line on inset) crosses the value of modulo $\pi/2$ at $\epsilon \rightarrow 0$, whereas
$\delta_{s}^{\rm SCAT(UP)}$ (dash line on inset) does not. Correspondingly, the cross section $\sigma_{s}^{\rm SCAT(CP)}$ develops a shape resonance at low $\epsilon$, in opposite to $\sigma_{s}^{\rm SCAT(UP)}$.
Note, the behavior of the phase shift  $\delta_{s}^{\rm SCAT(CP)}$ speaks to the fact that the binding potential of
 $e + {\rm Ne@\rm C_{60}}$ gets stronger when the polarizabilities of Ne and C$_{60}$ are coupled. This results in the emergence of an additional bound $s$-state in the field of $e + {\rm Ne@C_{60}}$ (compared
 to the approximation on uncoupled polarizabilities), thereby pushing $\delta_{s}^{\rm SCAT(CP)}$ to a larger value than that of $\delta_{s}^{\rm SCAT(UP)}$, at $\epsilon \rightarrow 0$. This is in accordance with
 the well known Levinson theorem: $\delta_{\ell} \rightarrow n_{\ell}\pi$ at $\epsilon \rightarrow 0$, $n_{\ell}$ being the total number of bound $\ell$-states in the system (valid when projectile-target electron exchange is neglected, as in the present work).

 Another interesting effect relevant to $s$-scattering is that
coupling between polarizabilities of C$_{60}$ and Ne  is found to largely cancel out the strong polarization impact of C$_{60}$ on $\sigma_{s}^{\rm Ne@C_{60}}$ in a broad energy region to the right of
 the maximum in  $\sigma_{s}^{\rm SCAT(CP)}$ (solid line). Indeed, there, $\sigma_{s}^{\rm SCAT(CP)}$ differs only somewhat from both $\sigma_{s}^{\rm HF}$ (dash-dot-dot line) and $\sigma_{s}^{\rm SCAT(A@)}$ (dash-dot line), but it differs significantly from $\sigma_{s}^{\rm SCAT(UP)}$ (dash line). We, thus, have unraveled an interesting effect. Namely, it is found that coupling of the large polarizability of C$_{60}$ with the considerably smaller polarizability of a compact encapsulated atom can have a significant impact on scattering of at least some of electronic partial waves off $A$@C$_{60}$. This, in turn, affects the total electron elastic scattering cross section $\sigma_{\rm tot}^{A@{\rm C_{60}}}$ as well, in the
 corresponding region of electron energies  (cp.\ dash and solid lines in the right bottom panel of Fig.~\ref{Ne}).

Next, we comment on one more interesting finding related to scattering of a $g$-partial electronic wave. The $g$-wave is seen to induce a strong narrow resonance in the $g$-partial scattering cross section $\sigma_{g}$ of
 $e + {\rm Ne@\rm C_{60}}$ scattering. The resonance is predicted by each of the utilized approximation: HF, SCAR(@A), SCAT(UP), and SCAT(CP). The origin of this resonance in $g$-scattering was established earlier in
  Refs.~\cite{DolmPRA15b,Winstead}, where it was shown that this is a shape resonance. One can see that
 the resonance in the cross section $\sigma_{g}$, calculated both in SCAT(UP) and SCAT(CP) approximations, emerges at noticeably lower energies and becomes significantly narrower than that in $\sigma_{g}$
 calculated in the framework of HF or SCAT(@A). The emphasized difference is interesting. The fact that the $g$-resonance emerges at the lowest energy and is the most narrow when calculated in the SCAT(CP) framework
 means that coupling between polarizabilities of C$_{60}$ and $A$ can result in a much stronger binding potential of $A$@C$_{60}$ than that calculated otherwise.

Lastly, the performed calculations demonstrate that accounting for polarization of $A$@C$_{60}$ by a scattered electron results in the emergence of near-zero minima in all partial cross sections $\sigma_{\ell}$s.
We refer to these minima as the Ramsauer minima by analogy with the known Ramsauer minima in electron elastic-scattering cross sections of some of free atoms (e.g., Xe). The presence of Ramsauer minima in all $\sigma_{\ell}$s for  $e + {\rm Ne@\rm C_{60}}$ scattering might prompt one to conclude that the total scattering cross section
$\sigma_{\rm tot}^{\rm Ne@C_{60}}$ could be so small that it might fall below the electron elastic-scattering cross section of free Ne, $\sigma_{\rm tot}^{\rm Ne}$, at some electron energy. This, however, does not happen, because
Ramsauer minima in the partial cross sections $\sigma_{\ell}^{\rm Ne@\rm C_{60}}$  for different $\ell$s emerge at somewhat different energies. Therefore, the minimum value of the total cross section
$\sigma_{\rm tot}^{\rm Ne@C_{60}}$ remains far greater than the cross section $\sigma_{\rm tot}^{\rm Ne}$. In the present case,  $\left.\sigma_{\rm tot}^{\rm Ne@C_{60}}\right|_{\rm min} \approx 60$~a.u.\
at $\epsilon \approx 1.4$ eV, whereas the electron elastic-scattering cross section of free Ne is one order of magnitude smaller, $\sigma_{\rm tot}^{\rm Ne} \approx 6$~a.u., at $1.4$ eV (cp.\ a solid or a dash line with a dot line in right bottom panel of Fig.~\ref{Ne}).

\subsection{$e + {\rm Xe@C_{60}}$ scattering}

\begin{figure}
\center{\includegraphics[width=\columnwidth]{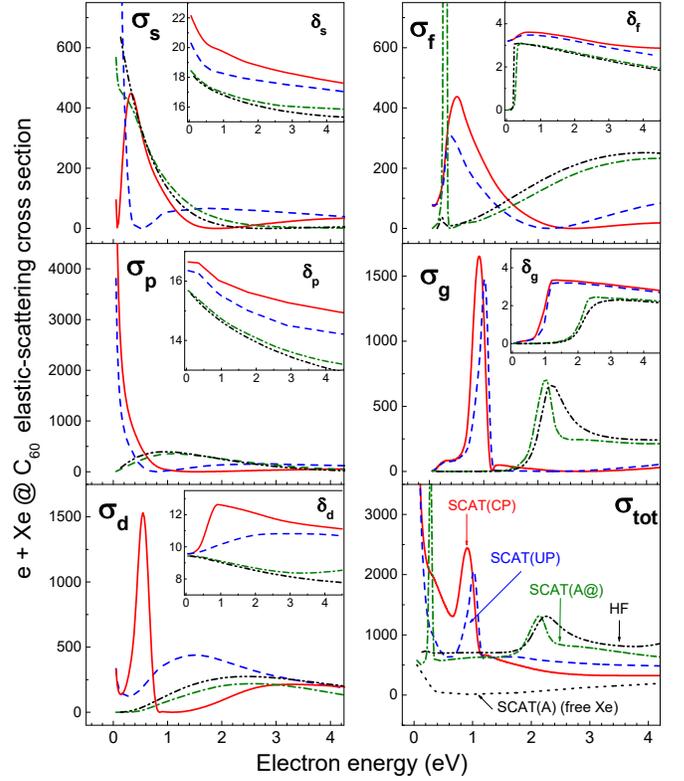}}
\caption{(Color online) Main panels: Calculated partial $\sigma_{\ell}(\epsilon)$ and total $\sigma_{\rm el}(\epsilon)$ cross sections (in atomic units)
for electron elastic scattering off Xe@C$_{60}$. Insets: Real parts $\delta_{\ell}(\epsilon)$ of the phase shifts (in units of radian); $\mu_{\ell} =0$ in this energy region.
The used styles of the plotted lines mark results obtained
in different approximation utilized in the present work, as follows:  dash-dot-dot, HF; dash-dot, SCAT(A@); dots, SCAT(A) (free Xe); dash, SCAT(UP); solid, SCAT(CP) (the most complete approximation).}
\label{Xe}
\end{figure}

We next explore changes in $e + A@{\rm C_{60}}$ scattering with increasing size and polarizability of an encapsulated atom $A$ which, however, is still a relatively compact atom of moderate polarizability.
To meet the goal, we study ${e + \rm Xe@C_{60}}$ scattering. Corresponding calculated data are displayed
in Fig.~\ref{Xe}.
%

Overall important similarities in, and difference between ${e + \rm Xe@C_{60}}$ and ${e + \rm Ne@C_{60}}$ scattering surface upon comparison of Fig.~\ref{Xe} with Fig.~\ref{Ne}.

We start focusing on similarities in ${e + \rm Xe@C_{60}}$ and ${e + \rm Ne@C_{60}}$ scattering. The overall impact of polarization of $\rm Xe@C_{60}$ by
an incident electron on ${e + \rm Xe@C_{60}}$ scattering remains significant. It results in the corresponding partial and total cross sections that differ strongly from those calculated without regard for the polarization
effect. Next, similar to $s$-scattering off $\rm Ne@C_{60}$,
the effect of coupled polarizabilities of Xe and C$_{60}$ [SCAT(CP) approximation] impacts strongly the $s$-scattering cross section. Indeed, this impact results in a strong shape resonance in $\sigma_{s}^{\rm SCAT(CP)}$ (solid line)
below $1$ eV of the electron energy. Furthermore, when coupling between polarizabilities of Xe and C$_{60}$ is accounted in the calculation of $s$-scattering, it largely cancels out the overall polarization impact of
$\rm Xe@C_{60}$ on $s$-scattering in a broad energy region to the right of the maximum in  $\sigma_{s}^{\rm SCAT(CP)}$ (solid line). Indeed, there, the $s$-cross section becomes about the same as the one calculated in the SCAT(A@) and HF approximations (dash-dot and dash-dot-dot lines). The emergence of the noted cancelation effect in two independent calculations -- $s$-scattering off Xe@C$_{60}$ and off Ne@C$_{60}$ -- speaks to the fact that the discovered effect is not accidental.  Next, coupling between polarizabilities of Xe and C$_{60}$ results in the emergence of Ramsauer minima in corresponding $\sigma_{\ell}$s. Lastly,
 similar to ${e + \rm Ne@C_{60}}$ scattering, the total cross section $\sigma_{\rm tot}^{\rm Xe@C_{60}}$ exceeds noticeably that for electron scattering
off free Xe (note, the cross section of electron elastic scattering off free Xe has its own low-energy Ramsauer minimum as well, in contrast to that for Ne).

The most striking differences between ${e + \rm Xe@C_{60}}$ and ${e + \rm Ne@C_{60}}$ scattering consist in the following. First, in contrast to a $d$-wave scattering off Ne@C$_{60}$, the $d$-wave scattering off Xe@C$_{60}$ is subject to a
particularly strong impact of coupled polarizabilities of Xe and C$_{60}$. Indeed, in the present case, the $d$-scattering cross section $\sigma_{d}^{\rm SCAT(CP}$ (solid line) develops the intense shape resonance below of $1$ eV and differs qualitatively and quantitatively from calculated data obtained in the frameworks of the SCAT(UP) (dash line) and other utilized approximations. Second, the overall
differences between calculated SCAT(UP) and SCAT(CP) electron scattering cross sections (both partial and total) are noticeably stronger in the case of ${e + \rm Xe@C_{60}}$ scattering
than in ${e + \rm Ne@C_{60}}$ scattering.

In summary, the highlighted differences between ${e + \rm Xe@C_{60}}$ and ${e + \rm Ne@C_{60}}$ underpin the effect of a bigger size and greater polarizability of a compact encapsulated atom $A$ on  $e+A@{\rm C_{60}}$ scattering.

\subsection{$e + {\rm Ba@C_{60}}$ scattering}

\begin{figure}
\center{\includegraphics[width=\columnwidth]{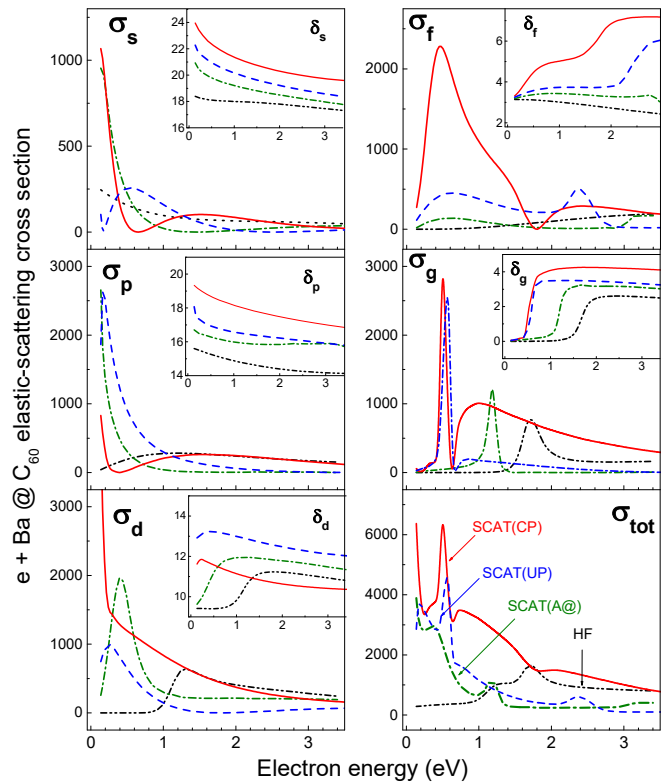}}
\caption{(Color online) Main panels: Calculated partial $\sigma_{\ell}(\epsilon)$ and total $\sigma_{\rm el}(\epsilon)$ cross sections (in atomic units)
for electron elastic scattering off Ba@C$_{60}$. Insets: Real parts $\delta_{\ell}(\epsilon)$ of the phase shifts (in units of radian); $\mu_{\ell} =0$ in this energy region.
The used styles of the plotted lines mark results obtained
in different approximation utilized in the present work, as follows:  dash-dot-dot, HF; dash-dot, SCAT(A@); dash, SCAT(UP); solid, SCAT(CP) (the most complete approximation).}
\label{Ba}
\end{figure}

In the earlier study of ${e + \rm Ba@C_{60}}$ scattering \cite{DolmPRA15b}, the impact of the polarization of \textit{only encapsulated Ba} on the scattering
process was accounted in the calculations [the SCAT(A@) approximation]. This impact was found to be considerable, due to a large value of the Ba polarizability.
The ${e + \rm Ba@C_{60}}$ scattering, therefore, stands out of the cases of ${e + \rm Ne@C_{60}}$ and ${e + \rm Xe@C_{60}}$ scattering. It sheds light
on features of the impact of coupled polarizabilities of a highly-polarizable atom and highly-polarizable C$_{60}$ on $e + A@{\rm C_{60}}$ scattering. Corresponding calculated data for
$e + {\rm Ba@C_{60}}$ scattering are displayed
in Fig.~\ref{Ba}.
%

An interesting feature, which caught our attention in the first head, is a cancelation effect of a new quality in the scattering process, compared to electron scattering off Ne@C$_{60}$ and off Xe@C$_{60}$ cases. Namely, the
account for the effect of coupled polarizabilities of Ba and C$_{60}$ in the calculation has practically annihilated the overall polarization impact of Ba@C$_{60}$ on the $s$-, $p$-, and $d$-partial cross sections
 above approximately $1.4$ eV of the electron energy. This follows clearly from the comparison of calculated $\sigma_{\ell}^{\rm SCAT(CP)}$ (solid lines) and $\sigma_{\ell}^{\rm HF}$ (dash-dot-dot lines) for $\ell = s$, $p$, and $d$. There, indeed,
$\sigma_{\ell}^{\rm SCAT(CP)} \approx \sigma_{\ell}^{\rm HF}$. For a $p$-cross section, this is true even for energies down to approximately $0.4$ eV, to a good approximation. For higher $\ell$s, especially for the $g$-cross section, as well as for the total cross section
 $\sigma_{\rm tot}$ above about $1.6$ eV, calculated SCAT(CP) data are noticeably closer to HF data than to cross sections calculated by accounting for polarization of only Ba (dash-dot line) or C$_{60}$
 (dash line).

 The reader can spot a number of other impressive differences between $\sigma_{\ell}$s, or between $\sigma_{\rm tot}$s for ${e + \rm Ba@C_{60}}$ scattering calculated in each of the utilized approximations,
 and/or between calculated data for ${e + \rm Ba@C_{60}}$ scattering, on the one hand, and ${e + \rm Ne@C_{60}}$ and ${e + \rm Xe@C_{60}}$ scattering, on the other hand.

In summary, similar to the earlier commentary on the differences between electron scattering off Ne@C$_{60}$ and Xe@C$_{60}$, the highlighted features of electron scattering off Ba@C$_{60}$
underpin, on a bigger scale, the effect of a greater size and polarizability of a compact encapsulated atom $A$ on  $e+A@{\rm C_{60}}$ scattering.

\section{Conclusion}

The present work provides the wealth of information about the role and significance of various polarization effects in $e +A@{\rm C_{60}}$ scattering unraveled in the framework
of reasonably simple approximations developed and utilized in the present work.

 As of today, this study provides the most complete
information about features of $e +A@{\rm C_{60}}$ scattering brought about by various confinement-related, static-related, and dynamical-related impacts of individual and coupled ``members'' of
$A$@C$_{60}$ on electron elastic scattering off this complex target. Each of these impacts has been found to bring specific features into $e +A@{\rm C_{60}}$ scattering. Spectacular effects in the scattering process, primarily associated with
polarization of $A$@C$_{60}$ by an incident electron have been unraveled, scrutinized, and fully detailed in this paper in a physically transparent manner.

The overall conclusion is that
 not only the individual impacts of uncoupled polarizabilities of the encapsulated atom $A$ (especially of a highly-polarizable atom) and C$_{60}$ on $e + A@{\rm C_{60}}$ scattering are generally important,
  but the correction term to $e + A@{\rm C_{60}}$ scattering, brought about by coupled polarizabilities of the atom $A$ and C$_{60}$, is essential. The significance
 of its impact on $e + A@{\rm C_{60}}$ scattering is found to increase with increasing size and polarizability of the atom $A$. For example, the effect of coupled polarizabilities of Ba and C$_{60}$ was demonstrated to result in considerable quantitative and qualitative modifications of partial cross sections for $e +{\rm Ba@C_{60}}$ scattering at lower energies, whereas it was found to largely annihilate the whole polarization impact on scattering of $s$, $p$, $d$, and $f$-partial waves above $\epsilon \approx 1.4 eV$.

The authors
hope that the present work will prompt other theorists to perform more rigorous calculations of $e +A@{\rm C_{60}}$ scattering. However, in the absence of experimental data on the subject, the question of which
theory is most appropriate to meet the goal would remain an open one. This should prompt experimentalists to perform corresponding measurements as well.

\section{Acknowledgements}

V.K.D. acknowledges the support by NSF Grant No.\ PHY-1305085.

\section*{References}


\begin{thebibliography}{}
%
\bibitem{DolmJPB} V. K. Dolmatov, M. B. Cooper, and M. E. Hunter, Electron elastic scattering off endohedral fullerenes $A$@C$_{60}$: The initial insight,
J. Phys. B \textbf{47}, 115002 (2014). DOI: 10.1088/0953-4075/47/11/115002
%
\bibitem{DolmPRA15a} V. K. Dolmatov, C. Bayens, M. B. Cooper, and M. E. Hunter, Electron elastic scattering and low-frequency bremsstrahlung on $A$@C$_{60}$: A model static-exchange approximation,
Phys. Rev. A \textbf{91}, 062703 (2015). DOI: 10.1103/PhysRevA.91.062703
%
\bibitem{AmusiaJETPL15} M. Ya. Amusia and L. V. Chernysheva, On the behavior of scattering phases in collisions of electrons with multi-atomic objects, JETP Lett. \textbf{100}, 503 (2015).
DOI: 10.1134/S0021364014210024
%
\bibitem{DolmPRA15b} V. K. Dolmatov, M. Ya. Amusia, and L. V. Chernysheva, Electron elastic scattering off $A$@C$_{60}$: The role of atomic polarization under confinement, Phys. Rev. A
\textbf{92}, 042709 (2015). 10.1103/PhysRevA.92.042709
%
\bibitem{AmusiaJETPL16} M. Ya. Amusia and L. V. Chernysheva, The role of fullerene shell upon stuffed atom polarization potential, JETP Lett. \textbf{103}, 286 (2016).
DOI: 10.7868/S0370274X16040093
%
\bibitem{Abrikosov} A. A. Abrikosov, L. P. Gorkov, and I. E. Dzyaloshinski, \textit{Methods
of Quantum Field Theory in Statistical Physics}, edited by R. A.
Silverman (Prentice-Hall, Englewood Cliffs, NJ, 1963).
%
\bibitem{ATOM} M. Ya. Amusia and L. V. Chernysheva, \textit{Computation of Atomic
Processes: A Handbook for the ATOM Programs} (IOP, Bristol, 1997).
%
\bibitem{JPCVKDSTM99} J.-P. Connerade, V. K. Dolmatov, and S. T. Manson, A unique situation for an endohedral metallofullerene, J. Phys. B \textbf{32}, L395 (1999).
DOI: 10.1088/0953-4075/32/14/108
%
\bibitem{DolmJPCS12} V. K. Dolmatov and D. A. Keating, Xe $4d$ photoionization in Xe@C$_{60}$, Xe@C$_{240}$, and Xe@C$_{60}$@C$_{240}$, J. Phys.: Conf. Ser. \textbf{388}, 022097 (2012).
 10.1088/1742-6596/388/2/022097
%
\bibitem{Gorczyca12} T. W. Gorczyca, M. F. Hasoglu, and S. T. Manson, Photoionization of endohedral atoms using R-matrix methods: Application to Xe@C$_{60}$, Phys. Rev. A \textbf{86}, 033204 (2012).
DOI: 10.1103/PhysRevA.86.033204
%
\bibitem{O'Sullivan13} Bowen Li, Gerry O'Sullivan, and Chenzhong Dong, Relativistic R-matrix calculation photoionization cross section of Xe and Xe@C$_{60}$,
J. Phys. B \textbf{46}, 155203 (2013). DOI: 10.1088/0953-4075/46/15/155203
%
\bibitem{Winstead} C. Winstead and V. McKoy, Elastic electron scattering by
fullerene, C$_{60}$, Phys. Rev. A \textbf{73}, 012711 (2006).  DOI: 10.1103/PhysRevA.73.012711
%
\bibitem{DolmJPCS15} V. K. Dolmatov, M. B. Cooper, and M. E. Hunter, $e + {\rm C}_{60}$ and $e + A@{\rm C}_{60}$ elastic scattering versus the parameters of the
C$_{60}$-model-square-well potential, J. Phys.: Conf. Ser. \textbf{635}, 112008 (2015). DOI: 10.1088/1742-6596/635/11/112008
%
\bibitem{Drukarev} G. F. Drukarev, \textit{Collisions of Electrons with Atoms and
Molecules, Physics of Atoms and Molecules}, Springer US, (1987), p. 252.
%
\bibitem{AmBaltPLA06} M. Ya. Amusia and A. S. Baltenkov, On the possibility of considering the fullerene shell C$_{60}$ as a conducting sphere, Phys. Lett. A \textbf{360}, 294 (2006).
DOI: 10.1016/j.physleta.2006.08.056
%
\bibitem{Safronova2010} J. Mitroy, M. S. Safronova, and C. W. Clark, Theory and applications of atomic and ionic polarizabilities, J. Phys. B \textbf{43}, 202001 (2010).
DOI:10.1088/0953-4075/43/20/202001

\end{thebibliography}
\end{document}